\documentclass[12pt]{article}
\usepackage{latexsym}
\usepackage{amsmath}
\usepackage{amssymb}
\usepackage[dvips]{graphicx}
\usepackage{latexsym}
\usepackage{enumerate,graphicx}
\usepackage[toc,page]{appendix}
\usepackage{setspace}
\usepackage{color,soul}
\newcommand{\K}{\mathbb{K}}

\def\be{\begin{equation}}
\def\ee{\end{equation}}

\def\wh{\widehat}

\def\p{\partial}

\def\ra{\rightarrow}

\def\h{\hat}

\def\mc{\mathcal}

\def\dsl{\displaystyle}

\def\a{\alpha}

\newtheorem{theo}{Theorem}[section]

\begin{document}
\title{
{\bf From the conformal self-duality equations to the Manakov-Santini system \vskip 20pt}}

\vskip 100pt
\author{
Prim Plansangkate\thanks{
  Email: prim.p@psu.ac.th} \\[15pt]
{\sl Applied Analysis Research Unit, Department of Mathematics and Statistics}, \\
 {\sl Prince of Songkla University, Hat Yai, Songkhla, 90110 Thailand}.}

\date{}
\maketitle

\vskip 50pt

\begin{abstract}
Under two separate symmetry assumptions, we demonstrate explicitly how the equations governing a general anti-self-dual conformal structure in four dimensions can be reduced to  
the Manakov-Santini system, which determines the three-dimensional Einstein-Weyl structure on the space of orbits of symmetry.  
The two symmetries  investigated are a non-null translation and a homothety, which are previously known to reduce the  second heavenly equation to the Laplace's equation and  the 
hyper-CR system, respectively. Reductions on  the anti-self-dual 
null-K\"ahler condition are also explored  in both cases.

\end{abstract}

\newpage
\setcounter{equation}{0}

\section{Introduction}

The integrability of the equations governing the anti-self-dual (ASD) conformal structures in four dimensions and the Einstein-Weyl (EW) structures in three dimensions is well known.
The twistor correspondences which reveal their integrability were given by Penrose \cite{P76} and Hitchin \cite{H82a}, respectively.  The relation between the two geometries was then established by
Jones and Tod \cite{JT85}, who proved that any ASD conformal structure with a non-null conformal  Killing symmetry
 gives rise to an EW space, and conversely given an EW space one can always find 
 an ASD spacetime with  a conformal Killing vector field  whose space of orbits is the EW space one started with.  
In fact the relation is not one-to-one; different ASD spacetimes can yield the same 
EW space, and starting from a given EW space one needs to solve the so-called generalised monopole equation, different solutions of which determine different ASD spacetimes.

In \cite{DFK15}, Dunajski {\it et al.} presented explicit forms of the equations for the ASD  and EW structures in specially adapted coordinate systems.  Using Cartan's approach they proved 
that any Lorentzian EW structure is locally determined by a solution of the Manakov-Santini system, which was first derived in \cite{MS06} as a generalisation of the 
dispersionless Kadomtsev-Petviashvili (dKP) equation.  They also proved that any ASD conformal structure in neutral signature can locally be written in terms of two functions  satisfying a system 
of two third order partial differential equations (PDEs).   We shall call this system, which was derived using the Pleba\'nski-Robinson coordinates, the ``ASD conformal equations".

According to the Jones-Tod  correspondence, one should be able to obtain the Manakov-Santini (MS) system from the ASD conformal equations under a conformal symmetry assumption.
The main aim of this paper is to demonstrate this explicitly.   We manage to do this in two cases of symmetry assumptions.  The first one is the reduction by a non-null translation and the second 
is by a homothetic Killing symmetry.  These two symmetries are previously known to reduce the second heavenly equation, determining the 
ASD Ricci-flat metrics, to the Laplace's equation \cite{FP79} and  the 
hyper-CR system \cite{D04}, respectively. 

 In both cases we also explore symmetry  reductions when the ASD conformal class admits a null-K\"ahler metric.
A metric of neutral signature  is said to be null-K\"ahler if it admits a covariantly constant real spinor. 
 It was  shown in \cite{D02} that
any  ASD null-K\"ahler 
metric  with a Killing symmetry preserving the parallel spinor  gives rise to an EW space with a parallel 
weighted vector field.  The latter is determined by the dKP equation  \cite{D01}, which is a reduction of the MS system.  
 In the case of the non-null translation, the translation is assumed to be a Killing symmetry of the ASD null-K\"ahler metric, representing the conformal class. Moreover,
it preserves the covariantly constant spinor.  Therefore one expects the ASD null-K\"ahler condition 
to give rise to the
dKP equation.  We demonstrate this explicitly.   On the other hand, the homothety is a conformal Killing symmetry, which rescales  the ASD null-K\"ahler metric to another metric, not necessarily 
null-K\"ahler.  Nevertheless we expected  to arrive at some reduction of the MS system, perhaps characterising the EW structure coming from an ASD conformal structure 
admitting a null-K\"ahler metric.    Instead, we find that the corresponding EW metric  is still determined by a general solution of the 
 the full MS system.

\vspace{0.3cm}

This paper is organised as follows.  In Section  \ref{sec:Review} the relevant  theorems which form the basis of our work  are reviewed.  
Then symmetry reductions by the non-null translation  are investigated   in Section \ref{sec:transl}.
There the ASD conformal equations  is  
   shown  to  reduce to the MS system under the symmetry assumption and  a Legendre transformation, and the dKP equation is derived directly from the ASD null-K\"ahler condition.
 Section  \ref{sec:hyperCR} explores reductions by  the homothetic Killing symmetry that reduces the second heavenly equation to the hyper-CR system. Assuming this conformal symmetry, 
we derive the MS system  from the ASD conformal equations, and show that even with the assumption of a null-K\"ahler metric in the ASD conformal class, the corresponding EW structure      is still governed by the MS system.
 Lastly, in Section \ref{sec:Bogdanov}, we discuss our attempts to explicitly realise the Bogdanov's system as another system of equations, other than the MS system, 
which describes a generic EW structure.

\vspace{1cm}


\section{Conformal self-duality and Einstein-Weyl equations}  \label{sec:Review}

Recall that a four-dimensional anti-self-dual (ASD) conformal structure $(M, [g])$ consists of a four-dimensional manifold $M$ and a conformal class of 
metrics $[g]$ whose Weyl tensor is anti-self-dual with respect to the Hodge star operator.    On the other hand, the Einstein-Weyl (EW) structure in three dimensions 
 $(\mc{W}, [h], D)$   consists of
 a three-dimensional manifold
$\mc{W},$ a conformal class of metrics $[h]$ and a torsion-free connection $D$ compatible with $[h],$ such that $D h = \nu \otimes h$ for some one-form $\nu,$ and the symmetrised Ricci 
tensor of $D$ is functionally proportional to $h.$  One can regard the EW  condition as equations for unknowns $h$ and $\nu,$ which determine the EW structure.

There are several explicit formulations for the ASD and EW conditions.  Here we shall focus on the 
explicit forms given in the main theorems of \cite{DFK15}, which are reviewed as Theorem \ref{ASDeq} and  \ref{MSeq} below.

\vspace{0.5cm}

\begin{theo} \label{ASDeq} {\rm \cite{DFK15} }
Any ASD conformal structure in signature $(2,2)$ can be  locally represented by a metric $g \in [g]$ of the form
\be \label{ASDmetric} g = dWdX + dZdY + F_Y dW^2 - (F_X+G_Y) dWdZ + G_X dZ^2, \ee
where $(W,X,Y,Z)$ are local coordinates on the manifold $M$ and the functions $F$ and $G$ satisfy a coupled system of third order PDEs,
\begin{eqnarray}
\p_X (Q(F)) - \p_Y (Q(G)) &=& 0, \label{asd1st} \\
(\p_W - F_Y \p_X + G_Y \p_Y) Q(G) + (\p_Z + F_X \p_X - G_X \p_Y) Q(F) &=& 0, \label{asd2nd}
\end{eqnarray}
with $Q$ being a second order differential operator given by 
\be \label{Q} Q = \p_W \p_X + \p_Z \p_Y - F_Y \p_X^2 - G_X \p_Y^2 + (F_X+G_Y) \p_X \p_Y.\ee
\end{theo}

\vspace{0.5cm}

\begin{theo} \label{MSeq} {\rm \cite{DFK15} }
 Any Lorentzian EW structure in three dimensions can be locally represented by a metric $h \in [h]$ and a one-form $\nu$ of the form
\be \label{hMS} h = -(dy -v_{x} dt)^2 + 4(dx - (u-v_{y})dt)dt,  \quad \nu = -v_{xx} dy + (4u_{x} - 2 v_{xy} + v_{x} v_{xx}) dt,  \ee
where $(x,y,t)$ are local coordinates on the manifold $\mc{W}$  and the functions $u$ and $v$ are solutions of the Manakov-Santini system
\be \label{MSsys}   P(u) + u_{x}^2 = 0, \qquad   P(v) = 0,  \ee
where 
\[ P =\p_{x} \p_{t} - \p_{y}^2 + (u - v_{y}) \p_{x}^2 + v_{x} \p_{x} \p_{y}. \]
\end{theo}

\vspace{0.3cm}

\noindent We note here that it was first shown in \cite{D08} that any solution of the Manakov-Santini (MS) system  (\ref{MSsys}) gives rise to an EW structure,
but it was proved in \cite{DFK15} that all EW structures arise in this way.

\vspace{1cm}

The relation between ASD conformal structures and the EW structures is given by the Jones-Tod correspondence  \cite{JT85}, which we shall state below for the ASD conformal structures in 
 signature $(2,2)$ and the Lorentzian  EW structures  using the notation of \cite{D02}.

\vspace{0.5cm}

\begin{theo} \label{JTcorresp} {\rm \cite{JT85, D02} }
Let $(M, [\h g])$ be a four-dimensional ASD conformal structure in neutral signature with a non-null conformal Killing vector field $K,$
and let $\mathcal{W}$ be the space of trajectories of $K.$  An EW structure on $\mc{W}$ is defined by the metric $h$ and the one-form $\nu$ given by 
\[ h= |K|^{-2} \h g - |K|^{-4} {\bf K } \odot {\bf K },   \qquad  \nu =  2\, |K|^{-2} *_{\h g} ({\bf K } \wedge d \, {\bf K}), \]
where $|K| = \h g_{ab} K^a K^b,$  ${\bf K}$ is the dual one-form of $K$ and $*_{\h g}$ is the Hodge star operator with respect to $\h g.$  

Conversely, suppose  $\mc{W}$ is a three-dimensional manifold  with a Lorentzian EW structure defined by a pair of metric and one-form $(h, \nu).$ Let  $V$ and 
$\a$ be a function of weight $-1$ and a one-form on $\mc{W},$  respectively.  If $V$ and $\a$ satisfy the generalised monopole equation
\be \label{genmoneq} *_h \left(d\, V + \frac{1}{2} \, \nu \, V\right) = d  \a, \ee
where $*_h$ is the Hodge star operator with respect to $h,$
  then
\be \label{ASDmetricfromJT}  g = V h - \frac{1}{V} (d \tau + \a)^2 \ee
is an ASD metric with a Killing symmetry generated by $\p_\tau.$
\end{theo}

\vspace{1cm}

From these three theorems, one expects that  under a non-null symmetry assumption the ASD conformal equations (\ref{asd1st}, \ref{asd2nd}) should  reduce to the MS system (\ref{MSsys}) or a 
special case of it.  In the next two sections we shall show that one can arrive at the MS system explicitly, using a symmetry assumption and
  a Legendre transformation.  The use of a Legendre transformation is motivated by the work of  Finley and  Pleba\'nski \cite{FP79}.

\vspace{1cm}


\section{Non-null translational symmetry reduction - \\ Lineage of the Laplace's equation }  \label{sec:transl}

A simplest symmetry to consider is a translation generated by a coordinate vector field.  
It turns out that such a symmetry is sufficient to reduce the ASD conformal equations (\ref{asd1st}, \ref{asd2nd})  to the full MS system (\ref{MSsys}), 
rather than a special case of it.  In particular, here we shall assume that  the ASD metric (\ref{ASDmetric}) admits the non-null translational Killing symmetry  generated by 
$\p_W$ (or equivalently $\p_Z$).  It was previously shown in the  aforementioned work of  Finley and  Pleba\'nski \cite{FP79} that  the second heavenly equation \cite{P75}, 
which governs the ASD Ricci-flat metrics, can be linearised to the three-dimensional Laplace's equation under this symmetry assumption.

\subsection{From ASD condition to the MS system}  \label{subsec:translMS}

Assume that the metric (\ref{ASDmetric}) admits a non-null translational Killing symmetry  generated by 
$\p_W.$  Without loss of generality, this implies  that the functions $F$ and $G$ in the metric (\ref{ASDmetric}) is independent of the $W$ coordinate.
For convenience, let  $\wh F:= Q(F)$ and $\wh G := Q(G),$ then equations (\ref{asd1st}) and (\ref{asd2nd}) become
\begin{eqnarray}
\wh F_X  - \wh G_Y &=& 0, \label{asdw1} \\
- F_Y \wh G_X  +  G_Y  \wh G_Y + \wh F_Z + F_X \wh F_X - G_X \wh F_Y &=& 0, \label{asdw2}
\end{eqnarray}
respectively, where
\begin{eqnarray}
  \wh F   &=& F_{ZY} - F_Y  F_{XX} - G_X F_{YY} + (F_X + G_Y)  F_{XY},  \label{Fh} \\
 \wh G   &=& G_{ZY} - F_Y  G_{XX} - G_X G_{YY} + (F_X + G_Y)  G_{XY}. \label{Gh}
\end{eqnarray}

Equation (\ref{asdw1}) implies the existence of a function $u(X,Y,Z)$ such that  $u_Y = \wh F$ and $u_X = \wh G.$  Working with the variable $u$, equation (\ref{asdw1}) 
is satisfied identically and  (\ref{asdw2}) becomes
\[ -F_Y u_{XX} +F_X u_{XY} + G_Y u_{XY} - G_X u_{YY} + u_{YZ} = 0. \]
The above equation can be written in terms of three-forms as 
\be \label{121} d F \wedge d u_X \wedge dZ \, + \, d u_Y \wedge (d G \wedge dZ  + dX \wedge dY) \, = \, 0. \ee

To facilitate the calculation, let us write the remaining two equations in terms of  three-forms.  Equations (\ref{Fh}) and (\ref{Gh}) are equivalent to
\begin{eqnarray}  - \, d u  \wedge dX \wedge dZ  & - & d F_Y \wedge d X  \wedge dY  \,+\,  F_Y \,d F_X \wedge d Y  \wedge dZ  \nonumber \\ 
 & -& G_X \, d F_Y \wedge d X  \wedge dZ \, - \, (F_X + G_Y) \, d F_Y \wedge d Y  \wedge dZ \, = \, 0,   \qquad  \label{Fhdiff}
\end{eqnarray}
and
\begin{eqnarray}  d u  \wedge dY \wedge dZ  & - & d G_Y \wedge d X  \wedge dY \,+\,  F_Y \,d G_X \wedge d Y  \wedge dZ  \nonumber \\
 & -& G_X \, d G_Y \wedge d X  \wedge dZ \, - \, (F_X + G_Y) \, d G_Y \wedge d Y  \wedge dZ= 0,   \qquad   \label{Ghdiff}
\end{eqnarray}
respectively.

\vspace{0.5cm}

Now we shall use  a Legendre transformation to show that the system (\ref{121}, \ref{Fhdiff}, \ref{Ghdiff}) is equivalent to the MS system (\ref{MSsys}). 
Note that one can assume that $F_Y \ne 0$ throughout a neighbourhood of  interest.  Otherwise,  the Killing vector field $\p_W$ would become null.
Introducing new independent variables  $(x,y,t),$ where $x=-F,$ $y = X$  and $t= -Z,$    we shall now write the system (\ref{121}, \ref{Fhdiff}, \ref{Ghdiff}) 
in $(x,y,t)$ coordinates as follows.

First, for equation (\ref{121}), one needs expressions of $u_X$ and $u_Y$ in  $(x,y,t)$ coordinates.
Suppose  $\h u(x,y,t)$ is a function defined by  $u(X,Y,Z) = \h  u(x(X, Y, Z), y(X), t(Z)).$
 Then by the Chain rule,
$ u_X = \h u_y + \h u_x\,  x_X$  and $u_Y = \h u_x \,  x_Y.$  Now, let $Y =\Phi(x, y, t),$ the derivatives  $x_X$ and $x_Y$ can be found by  differentiating
${Y = \Phi(x(X, Y, Z), y(X), t(Z))}$ with respect to $Y$ and $X,$ respectively.  This gives
$\dsl x_X = -\frac{\Phi_{y}}{\Phi_{x}}$  and $\dsl x_Y = \frac{1}{\Phi_{x}}.$  \, For convenience, from now we shall abuse the notation and drop the hat from $\h u,$ thus we have 
\[ u_X = u_y - \frac{u_x\,  \Phi_{y}}{\Phi_{x}} \quad \mbox{and} \quad  u_Y = \frac{u_x}{\Phi_x},   \]
where  $u$ on the right-hand side of each equation refers to the function $u$ expressed in the $(x, y, t)$ coordinates.  

Hence (\ref{121}) is  given  in $(x, y, t)$ coordinates  by
\[ d x \wedge d \left(u_y - \frac{u_x\,  \Phi_{y}}{\Phi_{x}} \right) \wedge dt \, + \, d \left(\frac{u_x}{\Phi_x}  \right) \wedge \big(d t \wedge d G  + dy \wedge d\Phi \big) \, = \, 0, \]
which is equivalent to
\be \label{122} -u_{yy} + u_{xt} +  \left(\frac{G_y -\Phi_t }{\Phi_x} \right) u_{xx} +  \left(\frac{\Phi_y - G_x}{\Phi_x} \right) u_{xy}
 + u_x \left( \p_x \Big(\frac{ G_y -\Phi_t }{\Phi_x} \Big)  + \p_y \Big(\frac{\Phi_y - G_x}{\Phi_x} \Big) \right)  = 0. \ee
Note that the condition $F_Y \ne  0$ guarantees that $dx \wedge dy \wedge dt \ne 0.$ \\

We shall now show that $(\ref{122})$ (which is (\ref{121}) in  $(x, y, t)$ coordinates) is equivalent to the first equation of the MS system (\ref{MSsys}) provided that (\ref{Fhdiff}) holds.  
Writing  (\ref{Fhdiff})  in  $(x, y, t)$ coordinates yields
\begin{eqnarray*} &  & u_x \,dx \wedge dy \wedge dt  \, + \, \frac{1}{\Phi_x} \, d \left( \frac{\Phi_y}{\Phi_x}\right) \wedge d \Phi  \wedge dt  \\
 & +& d \left( \frac{1}{\Phi_x}\right) \wedge  \left(  dy \wedge d \Phi - \frac{\Phi_y}{\Phi_x}\, d\Phi \wedge dt - \Big( G_y - \frac{G_x \Phi_y}{\Phi_x}  \Big)
 \, dy \wedge dt - \frac{G_x}{\Phi_x} \, d \Phi \wedge dt  \right) = 0,    
\end{eqnarray*}
where we have used   $\dsl x_X = -\frac{\Phi_{y}}{\Phi_{x}},$ $\dsl x_Y = \frac{1}{\Phi_{x}},$  $\dsl G_X = G_y -  G_x \frac{\Phi_y}{\Phi_x}$  
 and $\dsl G_Y = \frac{G_x}{\Phi_x}.$  

\vspace{0.3cm}

Since  $dx \wedge dy \wedge dt \ne 0,$ this implies
\be \label{up} u_x = \p_x \left(\frac{G_y -\Phi_t}{\Phi_x} \right)  + \p_y \left(\frac{\Phi_y - G_x}{\Phi_x} \right).\ee
Integrating (\ref{up}) with respect to $x$ gives
\[ u = \frac{G_y - \Phi_t}{\Phi_x} + \int   \p_y \left(\frac{\Phi_y - G_x}{\Phi_x} \right) \, dx.\]
As all functions involved are assumed to be real analytic and can thus be differentiated under the integral sign, we have  
\be \label{u} u =\frac{ G_y - \Phi_t }{\Phi_x} + v_y,  \quad \mbox{where} \quad  v=\int    \left(\frac{\Phi_y - G_x}{\Phi_x} \right) \, dx,
 \quad \mbox{and thus} \quad v_x = \frac{\Phi_y - G_x}{\Phi_x}.  \ee

With (\ref{up}) and (\ref{u}), equation (\ref{122}) becomes
\[   u_{xt} -u_{yy} +  (u-v_y)\, u_{xx} + v_x \, u_{xy} + u_x^2 = 0, \]
which is the first equation of (\ref{MSsys}).

\vspace{0.25cm}
 
The second equation of the MS system (\ref{MSsys}) arises from (\ref{Ghdiff}).  
In the coordinates $(x,y,t),$ (\ref{Ghdiff}) becomes
\begin{eqnarray*}  - \, u_{y}  \, + \, \frac{\Phi_y}{\Phi_x} u_{x}  & +& \frac{G_{xx}\Phi_t}{\Phi_x^2} \, - \, \frac{G_{x}\Phi_{xx}\Phi_t}{\Phi_x^3} \, - \, \frac{G_{xx}G_y}{\Phi_x^2} \, + \, \frac{G_x \Phi_{xx}G_y}{\Phi_x^3}  \\
&-&  \frac{G_{x}}{\Phi_x} \left( \frac{\Phi_y - G_x}{\Phi_x}\right)_y \, - \, \frac{G_{xy}\Phi_y}{\Phi_x^2} \, + \, \frac{G_{yy}}{\Phi_x}
\, - \, \frac{G_{xt}}{\Phi_x} \, + \, \frac{G_{x}\Phi_{xt}}{\Phi_x^2}   \,  = \, 0.    
\end{eqnarray*}  
With $u$ and $v_x$ given in (\ref{u}), the above equation can be rearranged to give
\[ v_{xt} - v_{yy} + (u-v_y) v_{xx} + v_x v_{xy} =0,\]
which is the second equation of (\ref{MSsys}).

\vspace{1cm}

The ASD metric (\ref{ASDmetric}) can be written in the form (\ref{ASDmetricfromJT}), where  
the three-dimensional EW metric $h$ is given by (\ref{hMS}) as follows.  First, let $w=W,$ in the new coordinates $(w, x, y, t),$  the metric (\ref{ASDmetric}) 
is given by 
\[ g = dw dy - d \Phi \, d t - \frac{1}{\Phi_x} dw^2 + \left( \frac{\Phi_y + G_x}{\Phi_x}\right) dw dt + \left( G_y -  \frac{\Phi_y G_x}{\Phi_x}\right) dt^2, \]
where we recall that $d Y = d \Phi(x, y, t)$ and the derivatives $F_X, F_Y, G_X, G_Y$ as functions of $(x,y,t)$ can be found using the chain rule.

\vspace{0.3cm}

Completing the square and using (\ref{u}) gives
\[ g =  \frac{\Phi_x}{4}  \Bigl(  \bigl( dy - v_x dt \bigr)^2 - 4 \bigl(dx - (u - v_y) dt \bigr) dt    \Bigr) 
- \frac{4}{\Phi_x} \left(    d \left( \frac{w}{2} \right) - \frac{\Phi_x}{4} \left(  \frac{\Phi_y + G_x}{\Phi_x} dt  + dy  \right)     \right)^2,  
  \]
which is of the form (\ref{ASDmetricfromJT}), where $\dsl V=  \frac{\Phi_x}{4},$ $\dsl \a = - \frac{\Phi_x}{4} \left(  \frac{\Phi_y + G_x}{\Phi_x} dt  + dy  \right) $ and the EW metric $h$ 
is given by
\[ h= \bigl( dy - v_x dt \bigr)^2 - 4 \bigl(dx - (u - v_y) dt \bigr) dt, \]
which is minus the metric given in (\ref{hMS}).
\vspace{1cm}

To end this section, let us note on what happens when  $F_Y = 0.$  It follows that the Killing vector field $K = \p_W$ becomes null.  Then the space of trajectories inherits a degenerate 
conformal structure and we no longer have an EW structure.  This situation was investigated by Dunajski and West \cite{DW07}.  Instead of the space of trajectories, they considered the 
two-dimensional space of ASD totally null surfaces in the manifold, called $\beta$-surfaces, containing the null Killing vector field, and showed that it admits a natural projective structure.
Local expressions for a general conformal structure admitting a null conformal Killing vector field were also derived in \cite{DW07}.  The form of a representative metric $g \in [g]$ was given 
in two cases; depending on whether the twist vanishes.  Recall that the twist is given by $\K \wedge d \K,$ where $\K := g(K, \cdot).$ \,  In the non-twisting  case, the ASD condition is 
completely solvable and the metric $g$ is given in terms of arbitrary functions of two variables.  If the twist is nonzero, then the ASD condition reduces to a linear PDE.

\vspace{0.2cm}

Let us now look back the ASD conformal equations (\ref{asdw1}) - (\ref{Gh}).  If $F_Y = 0,$ it follows 
from (\ref{Fh}) that $\wh F =0,$ and (\ref{asdw1}) and (\ref{asdw2}) reduce to one equation, $ \wh G_Y = 0.$
Together with (\ref{Gh}), this gives  

\vspace{-0.3cm}

\be \label{FY0eq} G_{ZYY}  - G_X G_{YYY}   + (F_X + G_Y) G_{XYY} = 0,  \ee
where $F$ is an arbitrary function of $X$ and $Z.$    Now, it can be shown that the twist $\K \wedge d \K$ vanishes if and only if $G_{YY} = 0.$

Hence in the non-twisting case, with $G_{YY} = 0,$ equation (\ref{FY0eq}) is satisfied trivally, and the metric (\ref{ASDmetric}) is given by 
\be \label{nontwistmetric}  g = dWdX + dZdY  - (A + B) dWdZ + (Y B_X + C) dZ^2, \ee
for arbitrary functions $A(X,Z),$ $B(X,Z)$  and $C(X,Z).$ 

In the twisting case, where $G_{YY} \ne 0,$ a theorem in \cite{DW07} guarantees that equation  (\ref{FY0eq}) can be linearised.  It is interesting to find an exact transformation which 
puts the metric (\ref{ASDmetric}) in the local form of \cite{DW07}  and read off the projective structure.  We leave this for future work.


\subsection{From ASD Null-K\"ahler condition to the dKP equation}  \label{subsec:translK}

In this section we shall assume that the ASD conformal class admits a null-K\"ahler metric.  Recall that 
a metric of signature  $(+ + - -)$ is called a null-K\"ahler metric if it admits a covariantly constant real spinor.  
 ASD null-K\"ahler metrics were studied in \cite{D02}.  In particular it was shown that all ASD null-K\"ahler metrics can locally  be written in the form 
\be \label{ASDnullKmetric} g = dWdX + dZdY - \theta_{YY} dW^2 + 2 \theta_{XY} dWdZ - \theta_{XX} dZ^2, \ee
where  function $\theta$ satisfies a fourth order PDE. 

Now the form (\ref{ASDnullKmetric}) can be obtained from the representative metric $g$ (\ref{ASDmetric}) in Theorem \ref{ASDeq} via 
the ansatz \cite{DFK15}
\be \label{FGnullK} F= -\theta_Y, \qquad G= -\theta_X. \ee
It also turns out that, via this ansatz,  the ASD conformal equations (\ref{asd1st}, \ref{asd2nd}) reduces to the fourth order PDE 
 for the null-K\"ahler condition.  To see this, note that with the ansatz  (\ref{FGnullK}),  $Q(F)$ and $Q(G)$ can be written as 
 \be \label{fdefine} Q(F) =  - \p_Y f, \quad  Q(G) =  - \p_X f, \quad \mbox{where} \quad f = \theta_{XW} + \theta_{YZ} + \theta_{XX} \theta_{YY} -  \theta_{XY}^2.\ee
Then one of the ASD conformal equations, (\ref{asd1st}), is then  satisfied  trivially by the commutativity of partial derivatives, and (\ref{asd2nd})
 becomes
\be \label{ansatzreduc} \wh Q(f) = 0, \quad \mbox{where} \quad \wh Q = \p_W \p_X + \p_Z \p_Y + \theta_{YY} \p_X^2 + \theta_{XX} \p_Y^2 - 2 \theta_{XY} \p_X \p_Y,  \ee
which is the fourth order equation in $\theta$ determining the ASD null-K\"ahler metric found in \cite{D02}.

As mentioned in the Introduction, any ASD null-K\"ahler metric with
symmetry preserving the parallel spinor determines an EW structure with a parallel weighted vector field, via  the Jones-Tod correspondence.  
The latter in turn is governed by a solution of the dKP equation \cite{D01}.     To be precise, we have

\begin{theo} \label{dKPthm} {\rm \cite{D01} }
 Any   three-dimensional  Lorentzian EW structure which admits a covariantly constant weighted vector field  can be locally represented by a metric $h \in [h]$ and a one-form $\nu$ of the form
\be  \label{hdKP} h = dy^2 - 4 dxdt  - 4 f  dt^2,  \qquad \nu = -4 f_x dt,  \ee
where the function $f(x,y,t)$ satisfies the dKP equation
\be \label{dKPeq}  (f_t - f f_x)_x = f_{yy}. \ee
\end{theo}

\vspace{0.3cm}

An explicit local expression of an ASD null-K\"ahler metric with symmetry in terms of solutions of the dKP equation 
and its linearisation (in `potential' form) is given in \cite{D02}.  The proof is based on the Jones-Tod correspondence and the twistor correspondences for the ASD  null-K\"ahler and the EW metrics.
   However, here we shall show  that one can obtain the dKP equation directly from the fourth order PDE (\ref{ansatzreduc}) using a non-null translational Killing symmetry and a Legendre transformation, 
in a similar way as in Section \ref{subsec:translMS}.

\vspace{0.5cm}

  Assume that the metric $g$ (\ref{ASDnullKmetric}) admits a Killing symmetry generated by $\p_W,$  which 
preserves the null-K\"ahler two form $dW \wedge dZ$  corresponding to the parallel spinor.
Then without loss of generality one can assume that $\theta$ is independent of $W,$ and  equation (\ref{ansatzreduc})
  can be written as a coupled system of equations as
\begin{eqnarray}
f_{YZ} + \theta_{YY} f_{XX} + \theta_{XX} f_{YY} -  2 \theta_{XY} f_{XY} &=& 0  \label{nullK1symeq} \\
f - (\theta_{YZ} + \theta_{XX} \theta_{YY} -  \theta_{XY}^2) &=& 0. \label{nullKsymf}
\end{eqnarray}

The system (\ref{nullK1symeq}, \ref{nullKsymf}) is equivalent to 
\begin{eqnarray}
d f_Y \wedge d X \wedge d Y  \, + \, d  f_X \wedge d \theta_Y \wedge d Z \, - \, d  f_Y \wedge d \theta_X \wedge d Z    &=& 0  \label{extnullK1symeq} \\
f \, d X \wedge d Y \wedge d Z \, - \, d \theta_X \wedge d \theta_Y \wedge d Z \, + \, d \theta_Y \wedge d Y \wedge d X  &=& 0. \label{extnullKsymf}
\end{eqnarray}

Now, introducing new variables $(x,y,t),$ where  $x = \theta_Y,$ $y=X,$ $t = - Z,$ and a new function $H(x,y,t) = \theta - x Y,$  
  we have
\[ d H = \theta_X d y -  Y d x - \theta_Z d t,  \]
where we have used $dy = dX$ and $dt = -dZ.$
This implies that
\be \label{coordrela} \theta_X = H_y,\quad Y = -H_x,   \quad \theta_Z = - H_t.  \ee
Making the substitution according to  (\ref{coordrela}),  equation (\ref{extnullKsymf}) gives
\be \label{fH} f =  \frac{H_{xt} - H_{yy} }{H_{xx}},  \ee
and  (\ref{extnullK1symeq}) becomes
\be \label{dKPdiff}  d \left( \frac{f_x}{H_{xx}} \right) \wedge d y \wedge d H_x  \, - \, d \left( f_y - \frac{f_x\,  H_{xy}}{H_{xx}} \right) \wedge d x \wedge d t \, - \,
 d \left( \frac{f_x}{H_{xx}} \right) \wedge d H_y \wedge d t    = 0,  \ee
where we have used $ \dsl f_Y = -\frac{f_x}{H_{xx}}$  and  $\dsl  f_X = f_y - \frac{f_x\,  H_{xy}}{H_{xx}}.$ \,  Then
making use of (\ref{fH}), equation (\ref{dKPdiff}) yields
the dKP equation (\ref{dKPeq}), as required.

\vspace{0.2cm}

Note that imposing the constraint $v=0$ and letting $u= -f$ \cite{DFK15} reduces the MS system (\ref{MSsys}) to the dKP equation (\ref{dKPeq}).

\vspace{1cm}
 To see that   the corresponding EW metric is of the form (\ref{hdKP}), first let $w=W$  and write 
the ASD null-K\"ahler metric (\ref{ASDnullKmetric}) in the $(w,x,y,t)$ coordinates as 
\[  g = dw dy + dt (H_{xy} dy + H_{xx} dx + H_{xt} dt) - \left(H_{yy} - \frac{H_{xy}^2}{H_{xx}}\right) dt^2 + 2  \frac{H_{xy}}{H_{xx}}\, dt dw + \frac{1}{H_{xx}} \,dw^2, \]
where we have used \,
$\dsl \theta_{YY}  = - \frac{1}{H_{xx}},$ 
$\dsl \theta_{XX}  = H_{yy} - \frac{H_{xy}^2}{H_{xx}}$ and
$\dsl  \theta_{XY}  = - \frac{H_{xy}}{H_{xx}}.$ \,

\vspace{0.3cm}

Using (\ref{fH}), the metric can be rearranged to be in the form
\be \label{ASDnullKmetricorbitform} 
g =  V \left( dy^2 - 4 dx dt - 4 f dt^2 \right) - \frac{1}{V} \, \left(d\left(\frac{w}{2}\right) + \alpha\right)^2,
\ee
where  $\dsl V = - \frac{H_{xx}}{4}$  and   $\a$ is a one-form given by $\dsl \a = \left(  \frac{H_{xy}}{2}\, dt  - V dy \right).$   Thus, comparing (\ref{ASDnullKmetricorbitform}) with (\ref{ASDmetricfromJT}), one sees that 
the EW metric on the space of orbits of symmetry is indeed of the form (\ref{hdKP}).
\vspace{1cm}

\noindent Let us now end this section by noting  that  a solution of the ASD null-K\"ahler equation (\ref{ansatzreduc}) is the trivial solution $f= 0,$ in which case  $\theta$ satisfies the second heavenly equation
\be \label{2ndheavenly} \theta_{XW} + \theta_{YZ} + \theta_{XX} \theta_{YY} - \theta_{XY}^2 = 0, \ee
and the metric is Ricci-flat.  The second heavenly equation
was first introduced by Pleba\'nski \cite{P75} as a form of the ASD Ricci-flat condition.   Any ASD Ricci-flat metric in the neutral signature can  be  written locally in the form of 
(\ref{ASDnullKmetric}), with $\theta$  a solution of the second heavenly equation (\ref{2ndheavenly}).  The idea of using the Legendre transformation in our calculation comes from the work of 
 Finley and Pleba\'nski \cite{FP79}.  There, it was shown that if an ASD Ricci-flat metric admits a non-null translational  Killing symmetry, then the metric can 
be put in the Gibbons-Hawking ansatz \cite{GH78}. Moreover, using a Legendre transformation, the second heavenly equation (\ref{2ndheavenly}) can be linearised to the three-dimensional Laplace's equation
\be \label{LaplaceRicciflat}  H_{yy} - H_{xt} = 0.\ee

\vspace{1cm}


\section{Homothetic symmetry reduction - Lineage of hyper-CR Einstein-Weyl equations}  \label{sec:hyperCR}
\setcounter{equation}{0}

In this section, we investigate the symmetry reduction by a particular homothetic Killing symmetry.  Our interest in this conformal Killing symmetry comes from the fact that 
it reduces the second heavenly equation - ASD Ricci-flat condition - to the so-called hyper-CR system, which is another reduction of the MS system.  

First, recall that the hyper-CR Einstein-Weyl structures are characterised by the absence of the derivative with respect  to the spectral parameter in the Einstein-Weyl Lax pair.  The name comes from
the fact that they arise on the space of orbits of a tri-holomorphic conformal Killing vector field  on a hyper-Hermitian manifold \cite{GT98} 
(see also references in physics literature such as \cite{P95, CTV96}).  Now, as the conformal 
Killing vector field preserves the hyperboloid of complex structures, these descend to a hyperboloid of Cauchy-Riemann structures on the EW space, thus the name hyper-CR. 
\, Via the Jones-Tod correspondence, a special solution to the generalised monopole equation (\ref{genmoneq}) can be chosen to construct a hyper-K\"ahler metric 
from a hyper-CR EW metric via (\ref{ASDmetricfromJT}).  On the other hand, the hyper-CR EW metrics were obtained as reductions of hyper-K\"ahler metrics  in  \cite{DT01}.

\vspace{0.3cm}

 Minor sign changes can be applied to obtain analogous results for Lorentzian signature.  It was shown in \cite{D04} that any 
 Lorentzian hyper-CR EW structure is locally determined by solutions to  
 a pair of quasi-linear PDEs  of hydrodynamic type \cite{FK04a, FK04b, MSh03, P03}
\be \label{hyperCRsys}
a_t + b_y + a b_x - b a_x = 0, \qquad \quad  a_y + b_x = 0,
\ee
which we shall call the hyper-CR system.  Via the symmetry reduction of a  pseudo-hyper-K\"ahler metric by a 
tri-holomorphic homothetic Killing vector field, it was shown that the second heavenly equation (\ref{2ndheavenly})  reduces to the hyper-CR system (\ref{hyperCRsys}), 
 and the EW metric $h$ and the associated one-form $\nu$ are locally of the form
\[
 h = (dy - a dt)^2 - 4(dx - a dy + b dt)dt,  \quad \nu = a_{x} dy + (a a_{x} + 2 a_{y}) dt.  
\]
Note  that 
(\ref{hyperCRsys}) is a reduction of the MS system under the constraint  $u=0$ \cite{DFK15} and setting $a=-v_x$ and $b=v_y.$

\vspace{0.3cm}

We shall now lift the hyper-K\"ahler, i.e. ASD Ricci-flat, condition and apply this symmetry reduction to a general ASD conformal structure.  It turns out that one also obtains the 
Manakov-Santini system, rather than a special case of it,  from this reduction.  Moreover, assuming this conformal symmetry on an ASD conformal class admitting a 
 null-K\"ahler metric, we recover the MS system from the fourth order PDE (\ref{ansatzreduc}).


\vspace{0.5cm}

\subsection{From ASD condition to the MS system} \label{subsec:hyperCRASD}

Without loss of generality, a
tri-holomorphic homothetic Killing vector field $K$ to a  pseudo-hyper-K\"ahler metric, of the form  (\ref{ASDnullKmetric}), is given by 
\be \label{hyperCRhomothethy} K = Z \frac{\p}{\p Z} + X \frac{\p}{\p X}.\ee
 Now, suppose a general ASD metric (\ref{ASDmetric})
\[g = dWdX + dZdY + F_Y dW^2 - (F_X+G_Y) dWdZ + G_X dZ^2\]
 admits a conformal symmetry 
generated by the homothety (\ref{hyperCRhomothethy}).  It follows that 
\be \label{hyperCRhom1} K(F_Y) = F_Y, \qquad  K(G_X) = - G_X, \qquad  K(F_X + G_Y) = 0.    \ee
With gauge freedom, it is possible to reduce (\ref{hyperCRhom1}) to 
\[ K(F) = F, \qquad  \quad K(G) =0,  \]
whose general solutions are given by
\[  F = Z \, P\left(\frac{X}{Z}, Y, W\right),  \qquad  \quad  G = R\left(\frac{X}{Z}, Y, W\right) \]
for arbitrary functions $P$ and $R.$

\vspace{0.3cm}

Following the change of variables used for the pseudo-hyper-K\"ahler case in \cite{D04},  let $\dsl U = \frac{X}{Z}$ and $S =\ln Z,$ then $\dsl K = \frac{\p}{\p S}.$
\, One can write the ASD conformal equations (\ref{asd1st}, \ref{asd2nd}) in the new variables $(U, Y, W)$ as follows.

\vspace{0.3cm}

First, let $\wh F := Q(F)$ and $\wh G := e^S Q(G),$ where $Q$ is the differential operator given by (\ref{Q}).  Notice that $\wh F$ and $\wh G$ depend only  on  $(U, Y, W),$
and equations (\ref{asd1st}) and (\ref{asd2nd}) become
\begin{eqnarray}
\wh F_U - \wh G_Y &=& 0, \label{hyperCR1st} \\
\wh G_W - P_Y \wh G_U + R_Y \wh G_Y +  (P_U-U) \wh F_U -  R_U \wh F_Y    &=& 0, \label{hyperCR2nd}
\end{eqnarray}
respectively.

Equation (\ref{hyperCR1st}) implies the existence of a function $u(U, Y, W)$ such that $\wh F = u_Y,$  $\wh G = u_U,$ and  (\ref{hyperCR1st})
is satisfied trivially.  Then, in terms of differential forms, (\ref{hyperCR2nd})  becomes
\be \label{hyperCR2nd1} 
d u_U \wedge dU \wedge dY - d u_U \wedge dP \wedge dW + d u_Y \wedge dR \wedge dW + U d u_U \wedge dU \wedge dW = 0.   
\ee
Moreover,  the definitions of $\wh F$ and $\wh G,$ which becomes
\begin{eqnarray*}
\wh  F &=& P_{UW} + P_Y - UP_{UY} - P_Y P_{UU} -R_U P_{YY} + (P_U + R_Y) P_{UY}  \label{hFPR} \\
\wh G &=&  R_{UW} - UR_{UY} - P_Y R_{UU} -R_U R_{YY} + (P_U + R_Y) R_{UY}, \label{hGPR}
\end{eqnarray*}
 yield
\begin{eqnarray}
d u \wedge dU \wedge dW \; =\; & &  d P_U \wedge dY \wedge dU  + d P \wedge dU \wedge dW  - U d P_U \wedge dU \wedge dW \nonumber \\
&-& d P \wedge dP_U \wedge dW + d R \wedge d P_Y \wedge dW, \label{hyperCRhF} \\
d u \wedge dY \wedge dW  \; =\; & &d R_U \wedge dU \wedge dY  -  U d R_U \wedge dW \wedge dU \nonumber \\
&+& d P \wedge dR_U \wedge dW - d R \wedge d R_Y \wedge dW,  \label{hyperCRhG}
\end{eqnarray}
respectively.

\vspace{0.3cm}

Now, using the change of variables and the Legendre transform similar to those in Section \ref{subsec:translMS}, the system
(\ref{hyperCR2nd1}, \ref{hyperCRhF}, \ref{hyperCRhG}) can be shown to be equivalent to the MS system (\ref{MSsys}) as follows.

First, assume that $R_U \ne 0$ throughout a neighbourhood of interest.  For if $R_U = 0,$ the conformal Killing vector field $K$ becomes null and the conformal structure on the 
space of orbits is degenerate.   Then introduce new variables $(x, y, t),$
where $x = -R(U, Y, W),$ $y=Y$ and $t= -W.$  Let $U = \Phi(x, y, t).$  In these new variables, (\ref{hyperCR2nd1}) becomes
\[ d \left(\frac{u_x}{\Phi_x}\right) \wedge \left( d \Phi \wedge dy + d P \wedge dt \right)
+ d \left( u_y - u_x \frac{\Phi_y}{\Phi_x}\right) \wedge d x \wedge dt - \Phi \, d \left(\frac{u_x}{\Phi_x}\right) \wedge d \Phi \wedge dt =0,  \]
which is equivalent to 
\begin{eqnarray}
 u_{xt} -u_{yy}  &+& \left( \frac{P_y - \Phi_t - \Phi \Phi_y}{\Phi_x} \right) u_{xx} +  \left(\Phi  + \frac{ \Phi_y - P_x}{\Phi_x} \right) u_{xy} \nonumber \\ 
&+&\left(  \p_x   \left( \frac{P_y - \Phi_t - \Phi \Phi_y}{\Phi_x} \right)  + \p_y \left( \Phi + \frac{ \Phi_y - P_x}{\Phi_x} \right)  \right) u_x = 0. \label{hyperCRMS1}
\end{eqnarray}
Note that $dx \wedge dy \wedge dt \ne 0$ is guaranteed by the assumption that $R_U \ne 0.$

\vspace{0.5cm}

Similarly, (\ref{hyperCRhG}) gives
\be \label{hyperCRux} u_x  =  \p_x   \left( \frac{P_y - \Phi_t - \Phi \Phi_y}{\Phi_x} \right)  + \p_y \left( \Phi + \frac{ \Phi_y - P_x}{\Phi_x} \right).  \ee
Integrating the above equation with respect to $x,$ we have 
\[u =  \frac{P_y - \Phi_t - \Phi \Phi_y}{\Phi_x} + \p_y \int  \left( \Phi + \frac{ \Phi_y - P_x}{\Phi_x} \right)  \, dx,\]
where we have used the assumption that all functions involved are real analytic and thus can be differentiated under the integral sign.  This implies that 
\be \label{hyperCRuvy}  u =  \frac{P_y - \Phi_t - \Phi \Phi_y}{\Phi_x} + v_y \ee
for a function $v(x,y,t)$ defined by
\be  \label{hyperCRv} v = \int  \left( \Phi + \frac{ \Phi_y - P_x}{\Phi_x} \right)  \, dx, \quad \mbox{and thus} \quad v_x =\Phi + \frac{ \Phi_y - P_x}{\Phi_x}. \ee

Making substitutions from  (\ref{hyperCRux})-(\ref{hyperCRv}), equation (\ref{hyperCRMS1}) becomes
\[ u_{xt}  -u_{yy} + (u-v_y) u_{xx} + v_x u_{xy} + u_x^2 = 0,\]
which is the first equation of the MS system (\ref{MSsys}). 

\vspace{0.3cm}

Lastly, through similar procedure, equation (\ref{hyperCRhF}) can be realised as  the second equation of the MS system, 
\[  v_{xt}  -v_{yy} + (u-v_y) v_{xx} + v_x v_{xy} = 0.\]

\vspace{1cm}

In the adapted coordinates $(S, U, Y, W),$ the metric (\ref{ASDmetric}) becomes
\[g = e^S \,  \big( \, dWdU + dS dY + P_Y dW^2 +(U-P_U - R_Y) dW dS +R_U dS^2 \, \big).\]

Now let $s=S,$ and with  the change of variables $y=Y,$ $t=-W,$ $x= -R(U, Y, W)$ and letting $U=\Phi(x,y,t)$ as before, the conformal metric
$\h g = e^{-S} g$ is given by
\[\h g = -dt d\Phi + ds dy + \left( P_y - \frac{P_x \Phi_y}{\Phi_x}\right) dt^2 + \left( \frac{P_x+\Phi_y}{\Phi_x} - \Phi \right) dt ds - \frac{1}{\Phi_x} ds^2.\]

This can be rearranged  in the form (\ref{ASDmetricfromJT})
\[\h g = V h - \frac{1}{V} (d \tau + \a)^2, \]

\vspace{-0.3cm}
\noindent where $\dsl \tau =\frac{s}{2},$ $\dsl V = \frac{\Phi_x}{4}$ and  $\a = (P_x + \Phi_y - \Phi \Phi_x) dt + \Phi_x dy.$  \, Then, 
using (\ref{hyperCRuvy}) and (\ref{hyperCRv}), one obtains  the EW metric $h$  precisely of the form (\ref{hMS}) albeit the minus sign.

\vspace{1cm}

Let us now comment on the case $R_U = 0$.  The homothety $K $ (\ref{hyperCRhomothethy}) is then null with respect to the conformal class, and the space of
$\beta$ surfaces admits a natural projective structure \cite{DW07}, as discussed at the end of Section \ref{subsec:translMS}.   One finds that the ASD conformal equations reduce to a single third order PDE
\be \label{RU0eq} L(P_{UU})=0, \quad \mbox{where} \quad L =\p_W + (P_U + R_Y -U) \p_Y - P_Y \p_U.\ee

The work of \cite{DW07} guarantees that  equation (\ref{RU0eq}) is completely solvable if the twist of $K$ vanishes, and linearisable if it does not.  In our coordinate system, 
it can be shown that the twist of $K$ is zero if and only if $P_{UU} = 1.$  Thus, in the non-twisting case it is clear that (\ref{RU0eq}) is satisfied trivially and the conformal class can be 
represented by a metric of the form (\ref{nontwistmetric}), upon renaming the coordinates. We leave the linearisation of (\ref{RU0eq}) in the twisting case for future work.

\vspace{0.5cm}


\subsection{Conformally Null-K\"ahler condition} \label{subsec:hyperCRK}

We shall now assume that the ASD conformal class $[g]$ admits a null-K\"ahler metric.  Thus $[g]$ can be represented by the  a null-K\"ahler metric $g$ of the form (\ref{ASDnullKmetric})
\[ g = dWdX + dZdY - \theta_{YY} dW^2 + 2 \theta_{XY} dWdZ - \theta_{XX} dZ^2, \]
where we recall that  $\theta$ satisfies a fourth order PDE, which can be written as a coupled system
\vspace{-0.3cm}
\begin{eqnarray}
f_{WX}+f_{YZ} + \theta_{YY} f_{XX} + \theta_{XX} f_{YY} -  2 \theta_{XY} f_{XY} &=& 0  \label{hyperCRnullK1} \\
f - (\theta_{XW} + \theta_{YZ} + \theta_{XX} \theta_{YY} -  \theta_{XY}^2) &=& 0. \label{hyperCRnullK2}
\end{eqnarray}

The assumption of  the homothetic symmetry generated by the vector field (\ref{hyperCRhomothethy}),  $\dsl K = Z \frac{\p}{\p Z} + X \frac{\p}{\p X},$ then implies that 
\[ K(\theta_{XX}) = - \theta_{XX}, \qquad K(\theta_{YY}) = \theta_{YY}, \qquad K(\theta_{XY}) = 0.\]
With gauge freedom, the system reduces to $K(\theta) = \theta,$ whose general solution is given by
\vspace{-0.3cm}
\[    \theta = Z \, P\left(\frac{X}{Z}, Y, W\right) \quad  \mbox{for an arbitrary function} \; P.  \]

Then, using the same  adapted coordinates $(S, U, Y, W)$   defined in Section \ref{subsec:hyperCRASD}, the system {(\ref{hyperCRnullK1}, \ref{hyperCRnullK2})}
becomes
\begin{eqnarray}
f_{UW} - U f_{UY} + P_{YY} f_{UU} + P_{UU} f_{YY} -  2 P_{UY} f_{UY} &=& 0 \label{hyperCRnullK11} \\
f - (P_{UW} + P_Y - UP_{UY} + P_{UU} P_{YY} -  P_{UY}^2) &=& 0. \label{hyperCRnullK21}
\end{eqnarray}

Following the same procedure as in Section \ref{subsec:translK}, we 
 introduce new variables $x=P_U,$\, $y=Y,$  \, $t=-W$  and a function ${H(x,y,t) = P-xU,}$ 
which satisfies  $\dsl H_t = -P_W,$ $\dsl H_y = P_Y$ and  $\dsl H_x = U.$  \,It turns out surprisingly that we recover the full MS system, 
rather than a reduction of it.  That is, the system {(\ref{hyperCRnullK11}, \ref{hyperCRnullK21})} is equivalent to the 
coupled system
\be  f_{yy} -  f_{xt}  + H_xf_{xy} - H_y f_{xx} +(f f_x)_x = 0 \label{hyperCRnullK12} \ee
\be H_{yy} - H_{xt} + H_x H_{xy} - H_y H_{xx} + f H_{xx} = 0,  \label{hyperCRnullK22} \ee
which is precisely the MS system (\ref{MSsys}) upon renaming $f = -u$ and $H=-v.$

\vspace{0.5cm}


\section{Discussion: Einstein-Weyl equations via Bogdanov's system}  \label{sec:Bogdanov}
\setcounter{equation}{0}

In this last section we discuss another local representation of the Einstein-Weyl equations, other than the MS system and its reductions.  It was suggested in \cite{DFK15} that a general solution of the Bogdanov's system should 
also  locally  determines a  generic EW structure.  It  is therefore natural to expect that the Bogdanov's system should arise from the ASD conformal equations  (\ref{asd1st}, \ref{asd2nd}) under a symmetry assumption. Our attempts to achieve  this goal have not been successful so far.  
 Nevertheless, let us present our exploration of a special case, namely the ASD  null-K\"ahler condition.

\vspace{0.2cm}

The Bogdanov's system \cite{B10} is given by
\be \label{Bogdanov} (e^{-\phi})_{tt} = m_t \phi_{xz} - m_x \phi_{zt}, \qquad m_{tt} e^{-\phi} = m_x m_{zt} - m_t m_{xz}. \ee
It is regarded as a two-component generalisation of the $SU(\infty)$-Toda equation: Setting $m=t,$ then the second equation is satisfied trivially and the first equation 
becomes the $SU(\infty)$-Toda equation 
\be \label{Todaeq} (e^{-\phi})_{tt} = \phi_{xz}.\ee

\vspace{0.2cm}

The reduction suggests a  Killing symmetry to be assumed on the ASD  conformal structure.  This is again motivated by the 
work of Finley and Pleba\'nski \cite{FP79}, which shows that under a certain Killing symmetry assumption the second heavenly equation, governing 
the ASD hyper-K\"ahler structure, 
reduces to  the $SU(\infty)$-Toda equation.  Therefore one  expects that assuming the same symmetry on an  ASD conformal  structure  
should give rise to  the Bogdanov's system or a special case of it.  

In what follows, we shall extend the reduction procedure in \cite{FP79}, to apply it to  an ASD  null-K\"ahler metric.
As a result,  a coupled system of equations governing the corresponding EW structure is obtained. 

\vspace{0.5cm}

Assume that a general ASD  null-K\"ahler metric (\ref{ASDnullKmetric}) admits a  Killing symmetry generated by the vector field
$K = W \p_W - X \p_X.$  Note that $K$ does not preserve the null-K\"ahler two form $dW \wedge dZ,$ corresponding to the parallel spinor.  Therefore we do not expect to recover the dKP equation.
 The  Killing equations are given by 
\[ K(\theta_{XX}) = 0, \qquad K(\theta_{XY}) = - \theta_{XY}, \qquad  K(\theta_{YY}) = - 2 \theta_{YY}, \]
which, using the gauge freedom, can be simplified  to 
\be \label{Killingeq} K( \theta) = -  2\theta.  \ee
The general solution of (\ref{Killingeq}) is
\be \label{theta}  \theta = W^{-2} \, P(WX, Y, Z).  \ee 
Let $V=WX,$ then the ASD null-K\"ahler condition {(\ref{hyperCRnullK1}, \ref{hyperCRnullK2})} becomes
\begin{eqnarray}
\h f - (V P_{VV} + P_{VV} P_{YY} - P_{VY}^2 - P_V + P_{YZ}) &=& 0 \label{nullKsym1} \\
V \h f_{VV} - \h f_V + \h f_{YZ} + P_{YY} \h f_{VV} + P_{VV} \h f_{YY} - 2 P_{VY} \h f_{VY} &=& 0,  \label{nullKsym2} 
\end{eqnarray}
where $\h f = W^2 \, f.$  Note that (\ref{nullKsym1}) implies that $\h f$ is a function  of $(V, Y, Z)$ only.

\vspace{0.5cm}

Equations (\ref{nullKsym1}) and (\ref{nullKsym2}) are equivalent to 
\be \label{extnullKsym1} \h f dY \wedge d V \wedge d Z  \, = \,  d  P_Y \wedge d P_V \wedge d Z \, + \, d  P_Y \wedge d Y \wedge d V  -  V d P_V \wedge   d Y \wedge d Z  
\, - \, P_V dY \wedge d V \wedge d Z,   \ee
\be \label{extnullKsym2} 
d P_Y \wedge  d \h f_V  \wedge d Z \, + \,  d \h f_Y \wedge d P_V \wedge d Z \, + \, d \h f_Y \wedge d Y \wedge d V 
\,-\, V d \h f_V \wedge d Y \wedge d Z 
\, - \, \h f_V   dY \wedge d V \wedge d Z   = 0, \ee
respectively.     Now, letting 
\be \label{gamma} \gamma = - d P_Y + P_V dZ - V dY,\ee equation (\ref{extnullKsym1}) becomes
\be \label{gdg} \gamma \wedge d \gamma = - \h f \, dY \wedge d V \wedge d Z.\ee

\vspace{1cm}

If $f$ and thus $\h f$  are identically zero, the metric (\ref{ASDnullKmetric}) is Ricci-flat and, as shown in  \cite{FP79}, is determined by 
the $SU(\infty)$-Toda equation.  To see this, note that when $f=0,$  (\ref{gdg}) becomes
\[ \gamma \wedge d \gamma = 0,\]
 where $\gamma$ is given by (\ref{gamma}).  By the Frobenius' theorem this implies the existence of functions 
$R, S$ such that 
\be \label{gFroK} \gamma = R \, d S.\ee

Equating (\ref{gFroK})   and (\ref{gamma}) and  introducing new coordinates $(t, x, z)$  where $t=V,$ $x=S(Y,V, Z)$ and  $z= -Z,$
it can be shown that 
\[ R_{tt} + (\ln R)_{xz} = 0,\]
which gives the $SU(\infty)$-Toda equation  (\ref{Todaeq}) 
upon setting  $R = e^{-\phi},$ as shown in \cite{FP79}.

\vspace{1cm}

For the general case where $f,$ and thus $\h f,$ are nowhere vanishing in some neighbourhood, equation (\ref{gdg}) and the Frobenuis' theorem imply  
the existence of functions $R, S, H$ of variables $(Y, V, Z)$ such that 
\be \label{PF} \gamma = d H +  R dS = - d P_Y + P_V dZ - V dY.\ee

Guided by the Ricci-flat case, we introduce new coordinates $(t, x, z),$  where $t=V,$ $x=S(Y,V, Z)$ and  $z= -Z,$ as before. 
Our aim is to transform the ASD  null-K\"ahler conditions (\ref{nullKsym1}, \ref{nullKsym2}), for  two unknowns $\h f$ and $P(V,Y,Z),$  
to a system of equations for $R(x,z, t)$ and $H(x,z,t).$

After some manipulation,  we find that   (\ref{nullKsym1}, \ref{nullKsym2})  reduces to a complicated system
of equation for $R$ and $H.$ That is,
\be \label{eq1} R_{tt} + \left( \frac{R_z - H_t R_t}{R+H_x}  \right)_x = 0, \ee
\be \label{eq2} A_t R_z - (B+ A_z) R_t - H_t B_x + (R+H_x) B_t  =0,  \ee
\[  \mbox{where} \quad A =    - \frac{\h f_x}{R_t}, \hspace{1cm} B =     \h f_t + \frac{\h f_x}{R_t} \left(\frac{R_z - H_t R_t}{R + H_x}\right),  \hspace{1cm}  \h f =   H_z - \frac{H_t R_z}{R_t}.  \]
However,  we have so far not been able to realise this system as the Bogdanov's system (\ref{Bogdanov}) or a   special case of it.

\vspace{0.5cm}

The corresponding EW metric in terms of solutions of  (\ref{eq1}, \ref{eq2}) is given by 
\be \label{Bogh} h =   \left( dt^2 + 4 H_t dt dz + 4 \frac{H_t R_z }{R_t} dz^2\right) + 4 (R+H_x) dx dz, \ee

A possible step forward could be to compare (\ref{Bogh}) with the  general expression for an EW metric \cite{FK14}  in terms of a 
solution to the Bogdanov's system (\ref{Bogdanov}), that is
\[ h = (m_x dx + m_t dt)^2 + 4 e^{-\phi} m_t  dx dz, \]

\vspace{-0.2cm}
\noindent and  realise a transformation between the two.  This problem still remains open.


\section*{Acknowledgements}

The author wishes to thank Maciej Dunajski for valuable comments and suggestions.  This
research is funded by the Thailand Research Fund, under the grant number TRG5880210.




\vspace{0.5cm}


\begin{thebibliography}{jafsdl}
\frenchspacing 

\bibitem{B10} Bogdanov, L. V. (2010) Non-Hamiltonian generalizations of the dispersionless 2DTL hierarchy, 
J. Phys. A {\bf 43}, 434008.




\bibitem{CTV96} Chave, T.,  Tod, K. P. and Valent, G. (1996) $(4,0)$ and $(4,4)$ sigma models with triholomorphic Killing vector, 
Phys. Lett. B {\bf 383}, 262-270.

\bibitem{D99} Dunajski, M. (1999) The twisted photon associated to hyper-Hermitian four manifolds, 
J. Geom. Phys. {\bf 30}(3), 266-281.

\bibitem{D01} Dunajski, M., Mason, L. J. and Tod, K. P. (2001) Einstein-Weyl geometry, the dKP equation and twistor theory, 
J. Geom. Phys. {\bf 37}, 63-92.

\bibitem{D02} Dunajski, M. (2002) Anti-self-dual four-manifolds with a parallel real spinor, 
Proc. R. Soc. A {\bf 458}, 1205-1222.


\bibitem{D04} Dunajski, M. (2004) A class of Einstein-Weyl spaces associated to an integrable system of hydrodynamic type, 
J. Geom. Phys. {\bf 51}, 126-137.

\bibitem{D08} Dunajski, M. (2008) An interpolating dispersionless integrable system, 
J. Phys. A {\bf 4}(31), 315202.





\bibitem{DFK15} Dunajski, M.,  Ferapontov, E. V. and Kruglikov, B. (2015) On the Einstein-Weyl and conformal self-duality equations,
J. Math. Phys. {\bf 56}, 083501.



\bibitem{DT01} Dunajski, M. and Tod, K. P.  (2001) Einstein-Weyl structures form hyper-Hermitian metrics with conformal Killing vectors, Diff. Geom. Appl. 
{\bf 14}, 39-55.

\bibitem{DW07} Dunajski, M. and West, S. (2007) Anti-self-dual conformal structures  with null Killing vectors from projective structures,  Comm. Math. Phys. {\bf 272}(1), 85-118.


\bibitem{FK04a} Ferapontov, E. V. and Khusnutdinova, K. R. (2004) On the integrability of  (2+1)-dimensional 
quasilinear systems, Comm. Math. Phys. {\bf 248}(1), 187-206.

\bibitem{FK04b} Ferapontov, E. V. and Khusnutdinova, K. R. (2004) The characterisation of two-component (2+1)-dimensional 
quasilinear systems, J. Phys. A: Math. Gen. {\bf 37}, 2949. 

\bibitem{FK14} Ferapontov, E. V. and Kruglikov, K. R. (2014) Dispersionless integrable systems in 3D and Einstein-Weyl geometry, J. Differ. Geom. 
{\bf 97}, 215-254. 

\bibitem{FP79}Finley, III, J. D., Pleba\'nski, J. F. (1979) The classification of all ${\mc H}$ spaces admitting a Killing vector,
J. Math. Phys. {\bf 20}(9), 1938-1945.


\bibitem{GT98} Gauduchon, P. and Tod, K. P.  (1998) Hyper-Hermitian metrics with symmetry, J. Geom. Phys. 
{\bf 25}, 291-304. 

\bibitem{GH78} Gibbons, G. W. and Hawking, S. W. (1978) Gravitational multi-instantons,
Phys. Lett. B {\bf 78}, 430-432.




\bibitem{H82a} Hitchin, N. J. (1982) Complex manifolds and Einstein’s equations, in {\it Twistor Geometry and Nonlinear Systems.} Berlin, Germany: Springer.



\bibitem{JT85} Jones, P. and Tod, K. P.  (1985) Minitwistor spaces and Einstein- Weyl spaces, Class. Quantum Grav.
{\bf 2}, 565-577. 



\bibitem{MS06} Manakov, S. V. and Santini, P. M. (2006) The Cauchy problem on the plane for the dispersionless Kadomtsev-Petviashvili equation, JETP Lett. {\bf 83}, 462-466.

\bibitem{MSh03} Martinez Alonso, L. and Shabat, A. B. (2003) Towards a theory of differential constraints of a hydrodynamic hierarchy, 
J. Nonlinear. Math. Phys. {\bf 10}(2), 229-242.

\bibitem{P95} Papadopoulos, G. (1995)  Elliptic monopoles and $(4,0)$-supersymmetry sigma models with torsion, Phys. Lett. B  {\bf 356}, 249-255.

\bibitem{P03} Pavlov, M. V. (2003) Integrable hydrodynamic chains,  
J. Math. Phys. {\bf 44}, 4134.

\bibitem{P76} Penrose, R. (1976) Nonlinear gravitons and curved twistor theory, Gen. Rel. Grav. {\bf 7}, 31-52.



\bibitem{P75} Pleba\'nski, J. F. (1975) Some solutions of complex Einstein equations,
J. Math. Phys. {\bf 16}, 2395-2402.











\end{thebibliography}
\end{document}